# NEW STUDY RESULTS THE SECONDARY SETTLEMENT FOR VERTICAL TOTAL STRESS ON HIGHWAY CONSTRUCTION BUILT ON SOFT GROUND IN THE MEKONG DELTA

**Hung Pham Van**
*Postgraduate, Master of Engineering*
*Sub-Institute of Transport Science and Technology in southern Vietnam*

**ABSTRACT:** Today, to calculate secondary settlement for vertical total stress by the analytic formula, we often use the formula of Raymond and wahls (1976)
This formula does not reflect that the secondary settlement for vertical total stress depends on the magnitude of the vertical total stress which impact sinkage in soft soil, and changes with depth and preconsolidation pressure. As a result, soft soil creep, resulted from the soft soil's vertical total stress, under road and highway at different depths will be different. The problems were solved in this article.

## 1. INTRODUCTION

To calculate secondary settlement for vertical total stress by the analytic formula, we often use the formula of Raymond and Wahls (1976) based on Casagrande's basic point of the secondary settlement for vertical total stress and the method of calculating the coefficient of secondary compression $C_\alpha$

$$S_s = \frac{C_t . H_1}{(1+e_1)} . (\log t_2 - \log t_1)$$

$$C_t = \frac{e_1 - e_2}{\log t_2 - \log t_1}$$

$C_t$ = secondary compression index

Is equivalant to

$$C_\alpha = \frac{C_t}{(1+e_1)}$$

$C_\alpha$ = coefficient of secondary compression

where, $H_1$ = the thickness of soil layer used to calculate creep settlement has the time t which the primary consolidation has $Q_t$ =100%.
$e_1$ = the void ratio at the end of primary consolidation and at the beginning of secondary settlement.
$e_2$ = the void ratio of soil at time $t_2$ during the period of secondary settlement.
$t_1$ = the time at the end of primary consolidation at some certain level of pressure.
$t_2$ = the certain time during the period of secondary settlement, we usually use times $t_1$ and $t_2$ which $t_1$ - $t_2$ equals to one cycle of logarithmic time.

## 2. CASAGRANDE'S POINT OF SECONDARY SETTLEMENT FOR VERTICAL TOTAL STRESS

Based on the research of soil sample carried out by an unconfined compression test and primary consolidation law, Casagrande pointed out:
- At pressure level of the experiment P, in Casagrande 's experimental curve, the primary consolidation ends before point E where $Q_t$ =100%) and from E onwards, it is the secondary settlement for vertical total stress.

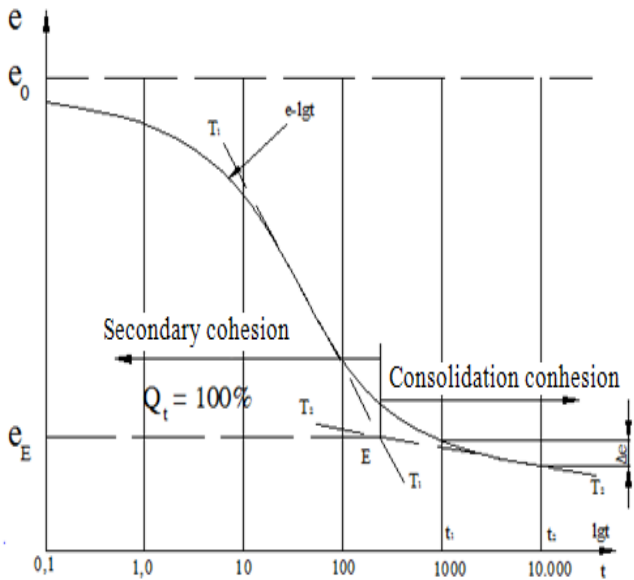

**Picture 1.** Compressibility and consolidation experimental results in semilogarithmic coordinates (e-lgt) corresponding to the level of pressure P

The secondary compression index:
$$C_t = \frac{e_1 - e_2}{\log t_2 - \log t_1}$$

The secondary compression coefficient:
$$C_\alpha = \frac{C_t}{(1 + e_1)}$$

$e_E$ = the void ratio at the end of primary consolidation.

Thus we see that according to Raymond and Wahls as well as Casagrande's basic point of the secondary settlement:
- At any level of pressure, the primary consolidation settlement ends and afterwards, it is the secondary settlement.
- If the earth mass load increases, the above primary consolidation settlement will repeat corresponding to the new load level.
- The experimental settlement curve ($s_t$, lg t) section where the secondary settlement occurs has an almost unchanged angle of slope ß.
- The creep settlement does not depend on the magnitude of each effective stress.

*The above results also have been used, cited and published in many books, journals of authors such as Braja M. Das, G. Mersri Phan Trường Phiệt.*

## 3. SETTING UP THE EQUATION OF THE VERTICAL CREEP STRAIN FOR VERTICAL TOTAL STRESS AS THIS ARTICLE'S AUTHOR SUGGESTS

According to N.M. Gerxevanov (1948), relation equation of plastic deformation at various levels of pressure is:

$$e_2 = e_1 - \frac{1}{A} \ln \frac{\sigma_{2C} - C}{\sigma_{1C}}$$

A is a material constant, C is a curve parameter.

$$\Rightarrow e_1 - e_2 = \frac{1}{A} \ln \frac{\sigma_{2C} - C}{\sigma_{1C}}$$

$$\Rightarrow \Delta e = e_1 - e_2 = \frac{C_t}{2,3} \frac{dt}{t} = \frac{1}{A} \ln \frac{\sigma_{2C} - C}{\sigma_{1C}}$$

According to Lomtadze and Gerxevanov, when $\sigma_{1C} = 1 kG/cm^2$, $c << \sigma_{1C}$. Hence, we could rewrite the above equation in the following form:

$$\frac{C_t}{2,3} \frac{dt}{t} = \frac{1}{A} \ln \frac{\sigma_{2C}}{\sigma_{1C}}$$

If:

$$\frac{\sigma_{2C}}{\sigma_{1C}} = 1 \Rightarrow \ln \frac{\sigma_{2C}}{\sigma_{1C}} = 0 \Rightarrow C_t = 0 \Rightarrow e = e_1$$

If:

$$\frac{\sigma_{2C}}{\sigma_{1C}} = 2,718 \Rightarrow \ln \frac{\sigma_{2C}}{\sigma_{1C}} = 1 \Rightarrow \frac{1}{A} = \frac{C_t}{2,3} \frac{dt}{t} = de$$

$$\Rightarrow \frac{1}{A} = de = e_1 - e_2 \Rightarrow A = \frac{1}{e_1 - e_2}$$

Consequently, considering $\frac{\sigma_{2C}}{\sigma_{1C}}$ in the range from 1 to 2,718, we have:

$$\frac{C_t}{2,3}\frac{dt}{t}=(e_1-e_2)\ln\frac{\sigma_{2C}}{\sigma_{1C}}=C_t(\log t_2-\log t_1)$$

$$\Rightarrow e_1-e_2=\frac{C_t}{\ln\dfrac{\sigma_{2C}}{\sigma_{1C}}}(\log t_2-\log t_1) \text{ If}$$

$$\frac{\sigma_{2C}}{\sigma_{1C}}=2{,}718\Rightarrow\ln\frac{\sigma_{2C}}{\sigma_{1C}}=1$$

We will obtain the relation equation according to Raymond and Wahls again

$$e_1-e_2=C_t(\log t_2-\log t_1)$$

Also, from:

$$(e_1-e_2)\ln\frac{\sigma_{2C}}{\sigma_{1C}}=\frac{C_t}{2,3}\frac{dt}{t}=\frac{C_t}{2,3}d(\ln t)$$

$$(e_1-e_2)\ln\frac{\sigma_{2C}}{\sigma_{1C}}=C_t d(\log t)=C_t(\log t_2-\log t_1)$$

$$\Rightarrow C_t=\frac{e_1-e_2}{\log t_2-\log t_1}\ln\frac{\sigma_{2C}}{\sigma_{1C}}$$

Thus, creep settlement of soil layer with thickness $H_1$ is calculated from the following formula:

$$S_s=\frac{\Delta e(t)}{1+e_1}H_1$$

$$\Delta e(t)=\frac{C_t}{\ln\dfrac{\sigma_{2C}}{\sigma_{1C}}}(\log t_2-\log t_1)$$

$$\Rightarrow S_s=\frac{C_t\cdot H_1}{(1+e_1)\ln\dfrac{\sigma_{2C}}{\sigma_{1C}}}\cdot(\log t_2-\log t_1)$$

Thus, based on N.M. Gevxevanov (1948) and Lomtadze's research results, we have proved that the equation of the vertical creep deformation for vertical total stress is consistent with Raymond & Wahls (1976)'s equation to calculate secondary settlement for vertical total stress, but yet more general and in special cases, Ph.D. Student's equation becomes the Raymond & Wahls's equation.

- The general creep settlement equation for vertical total stress in relationship with various levels of compression pressure is:

$$S_s=\frac{C_t\cdot H_1}{(1+e_1)\ln\dfrac{\sigma_{2C}}{\sigma_{1C}}}\cdot(\log t_2-\log t_1)$$

Where, $e_1$ = the void ratio at time $t_1$, the end of primary consolidation, corresponding to some compression pressure.

$H_1$ = the thickness of soil layer after the end of primary consolidation corresponding to some compression pressure.

$t_1$ = the time at the end of primary consolidation corresponds to the soil sample having the void ratio $e_1$ of the primary consolidation.

$t_2$ = time after the ending time $t_1$ of primary consolidation.

$e_2$ = the void ratio corresponding to the time $t_2$ of the creep period.

$C_t$ = the secondary compression index is calculated from the following formula:

$$C_t=\frac{e_1-e_2}{\log t_2-\log t_1}\cdot\ln\frac{\sigma_{2C}}{\sigma_{1C}}$$

$C_\alpha$ = the coefficient of secondary compression is calculated from the following formula:

$$C_\alpha=\frac{C_t}{(1+e_1)\ln\dfrac{\sigma_{2C}}{\sigma_{1C}}}$$

$\sigma_{1C}$ = levels of compression pressure causing the primary consolidation strain and creep is usually taken from the preconsolidated pressure.

$\sigma_{2C}$ = the desired compression pressure causing creep strain due to the impact of construction loads. The compression pressure $\sigma_2$ under structure foundation is chosen according to the average value of each layer in agreement with the effective normal stress diagram in active zone. Large creep strain will take place in the zone of higher compressive stress than preconsolidated compressive stress. $\sigma_{1C}\equiv\sigma_p$

$\ln\dfrac{\sigma_{2C}}{\sigma_{1C}}$ = dimensionless creep parameter.

We also notice that in the particular case:

$\dfrac{\sigma_{2C}}{\sigma_{1C}} = 2{,}718 \Rightarrow \ln \dfrac{\sigma_{2C}}{\sigma_{1C}} = 1$, $C_t$, $C_\alpha$, $S_s$ will have the same formular which Raymond and Wahls suggested in 1976:

$$C_t = \dfrac{e_1 - e_2}{\log t_2 - \log t_1}; \quad C_\alpha = \dfrac{C_t}{1+e_1}$$

$$S_s = \dfrac{C_t \cdot H_1}{(1+e_1)} \cdot (\log t_2 - \log t_1)$$

Here, this article's author notes that:
- Secondary settlement for the shear stress τ is different from the secondary settlement for vertical total stress which will be presented in another article.
- The soft soil under embankments when affected by the impact load will be layered into multiple layers. Complying with Casagrande's basic point which is that the primary consolidation settlement occurs first with the following creep deformation, some layers are in the process of primary consolidation settlement and some layers are in the process of creep deformation.
- The soft soil under embankments when affected by the impact load is always in the process of creep deformation due to shear stress τ when τ<τ$_{lim}$=σtgφ+c$_c$

τ$_{lim}$= Maslov's creep threshold

## 4. APPLICATION OF CALCULATION OF SECONDARY SETTLEMENT FOR VERTICAL TOTAL STRESS AT A PARTICULAR CONSTRUCTION IN THE MEKONG DELTA

The construction which this research applies is Tan Thanh's dike road, Go Cong town, Tien Giang Province.

a- Input data

| Highway width | 8.2 | m |
|---|---|---|
| Backfill height | 3.25 | m |
| Backfill talus slope | 2 | |
| Counterweight berm width | 0 | m |
| Counterweight berm height | 0 | m |
| Counterweight berm talus slope | 0 | |
| Distance from road centerline to calculate creep settlement | 0 | m |
| Settlement provision height | 4.17 | m |
| Empirical coefficient to calculate settlement | 1.4 | |

| Soil layer | γ$_{nature}$ | C$_c$ | e$_0$ | σp(kG/cm²) |
|---|---|---|---|---|
| 1 | 1.700 | 0.326 | 1.338 | 0.868 |
| 2 | 1.540 | 0.719 | 2.015 | 0.616 |
| 3 | 1.850 | 0.322 | 0.921 | 0.852 |
| 4 | 2.050 | 0.198 | 0.605 | 0.459 |

| C$_r$ | C$_\alpha$ characteristic | Depth (m) |
|---|---|---|
| 0.035 | | 1.600 |
| 0.070 | 0.0061 | 10.000 |
| 0.055 | | 11.000 |
| 0.053 | | 17.000 |

b- Calculation results

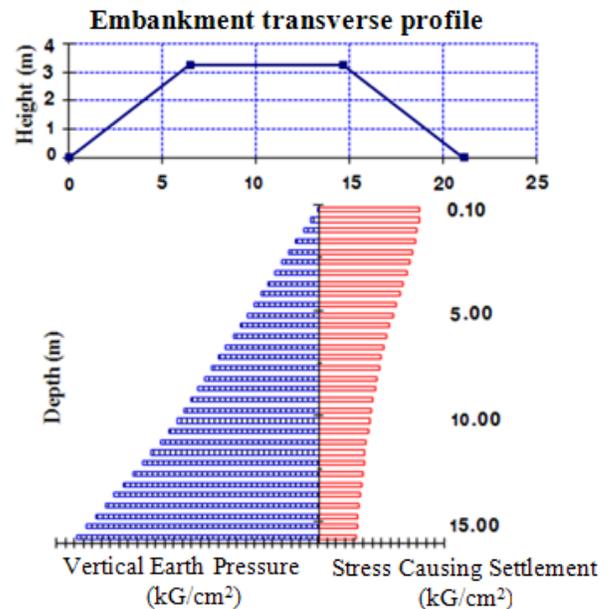

**Picture 2.** Embankment transverse profile and the depth of stress zone causing settlement

*Table 1. Comparison between creep settlement according to R&W and the formula proposed by this article's author at a point on the road centerline (creep settlement time is 40 years)*

| Thickness of soil layer | $C_\alpha$ Raymond & Wahls | $C_\alpha$ H.Pham-Van | $\sigma_p$ (kG/cm²) | $\sigma_2$ (kG/cm²) |
|---|---|---|---|---|
| 2 | 0.0022 | 0.0000 | 0.868 | 0.677 |
| 2 | 0.0108 | 0.0861 | 0.868 | 0.984 |
| 2 | 0.0018 | 0.0089 | 0.380 | 1.287 |
| 2 | 0.0018 | 0.0173 | 0.852 | 1.588 |
| 2 | 0.0018 | 0.0136 | 0.852 | 1.887 |
| 2 | 0.0003 | 0.0000 | 6.547 | 2.265 |
| 2 | 0.0003 | 0.0000 | 6.547 | 2.815 |
| 1.5 | 0.0013 | 0.0078 | 2.615 | 3.091 |

| $C_\alpha$ Characteristic Raymond & Wahls | $C_\alpha$ Characteristic H. Pham-Van | Creep settlement (cm) Raymond & Wahls | Creep settlement (cm) H. Pham-Van |
|---|---|---|---|
| 0.0061 | 0.0170 | 2.661 | 7.461 |

*Table 2. Synthesis results*

| The effective zone of settlement | 15.5 m |
|---|---|
| Primary consolidation settlement | 91.74 cm |

| Active zone: 15.5m | Consolidation settlement: 65.53cm |
|---|---|
| Vertical earth pressure: 2.7 kG/cm² | Instant settlement: 26.21 cm |
| Stress causing settlement: 0.42 kG/cm² | Total settlement: 91.74 cm |
| Creep settlement: 99.20 cm | |

**Notice:**

$C_C$ = compression index

$C_r$ = swell index

$C_\alpha$ = coefficient of secondary compression

$\gamma$ = moist density

$\sigma_p$ = preconsolidation pressure

$S_s$ = creep settlement

## 5. COMMENTS AND CONCLUSIONS

5.1. The average value of the secondary settlement for vertical total stress calculated using this article's author's method is 7,46cm, which is 2,75 times higher and closer to the Mekong Delta's reality than that of Raymond & Wahls' method, which is approximately 2,66cm in the period of 40 years.

5.2. The formula - suggested by this article's author - to calculate the secondary settlement for vertical total stress, which is more general and suitable to the Mekong Delta's soft ground, is proposed based on taking the representative value of vertical total stress of each thin soil layers, giving us a more precise result. In the special case, this general formula will return to the formula of Raymond & Wahls in 1976. This formula quantifies creep zone accurately according to the effective stress value, larger stress will result in larger creep strain.

## 6. REFERENCES

BraJa.M.DAS. Principles of foundation engineering. PWS. Kent Publishing Company (1987).

Braja.M.Das. Advanced soil mechanics. Taylor & Francis (1997).

J.H. Atkinson, PL. Bransby. The mechanics of soils, An introduction to critical state soil mechanics. Mc.Graw, Hill Book Company Limited (1982).

Ter. Soil behavior and critical state soil mechanics. Cambridge University press (1994).

Н.Н.Маслов, физико. техническАя тЕория ползучЕсти глинистых грунтов в прАктикЕ строительствА. МосквА строЙизДАт (1984).

V.D.Lomtadze. Địa chất công trình – Thạch luận công trình. NXB Đại Học và Trung Học Chuyên Nghiệp (1978).

V.D. Lomtadze. Địa chất công trình chuyên môn. NXB Đại Học & Trung Học Chuyên Nghiệp(1983).

V.D. Lomtadze. Địa chất động lực công trình. NXB Đại Học và Trung Học Chuyên Nghiệp (1982).

Phan Trường Phiệt. Cơ học đất và ứng dụng tính toán công trình trên nền đất theo trạng thái giới hạn. NXB Xây Dựng (2005).